\documentstyle[preprint,tighten,eqsecnum,epsfig,aps]{revtex}
\input{psfig}

\def\beq{\begin{equation}}   \def\eeq{
\end{equation}}

\begin{document}
\title{ Space-Time Evolution of
Ultrarelativistic Quantum Dipoles in  Quantum Electrodynamics.
  } \author{ B. Blok\thanks{E-mail: blok@physics.technion.ac.il} }
\address{Department of Physics, Technion -- Israel Institute of
Technology, Haifa 32000, Israel}
\maketitle

\thispagestyle{empty}

\begin{abstract}
We discuss space-time evolution of ultrarelativistic quantum
dipole in QED. We show that the space-time evolution can be
described, in a certain approximation, by means of a regularized
wave function, whose parameters are determined by the process of
the dipole creation by a local current. We derive using these wave
functions the dipole expansion law, that is found to coincide
parametrically in the leading order with the one suggested in ref.
\cite{Farrar}.

 \end{abstract}

 \setcounter{page}{1} \section{Introduction}
\par
The problem of a space-time time evolution of a
electromagnetic field surrounding electrically
charged quantum
particle has always been under investigation, starting from the
classical refs.  \cite{Chudakov,Feinberg,Gribov1} (see also
closely related research in classical electrodynamics
\cite{Bolotovskii}). In particular, Gribov stressed the connection
between the space-time evolution in quantum field theory and  the
parton model and deep-inelastic scattering. He argued  that at high
energies the space-time evolution is a well defined physical
concept, that directly follows from the study of the relevant
Feynman diagrams.
\par The problem of the space-time
evolution of a neutral dipole in gauge quantum field theory, i.e. QED or QCD,
recently attracted a lot of attention. This problem is more
complex, than the corresponding problem for a single charged
particle since one must take into account several effects, such as
field regeneration, charge screening and space-time evolution of
the dipole wave function, simultaneously.

\par  The  studies of the space-time evolution of the
electromagnetic dipoles started long time ago in connection to the
propagation of the fast $e^+-e^-$ pairs through matter. The latter
research, both experimental
\cite{Perkins} and theoretical \cite{Iekutieli}, led to significant
progress in the understanding of the properties of the dipoles
both in classical and in quantum electrodynamics.
\par More recently, a lot of interest was attracted to the space-time
evolution and screening in the quantum electromagnetic dipoles,
leading to the concept of the charge transparency \cite{FS13}.
This research is closely connected to the recent developments in
the study of deep inelastic scattering in QCD, where similar ideas
lead to the concept of the color transparency and the discovery
that a color neutral dipole is a correct degree of freedom for
deep-inelastic scattering (for the review and the proper
references see refs. \cite{FS1,FS,M}). For the sufficiently small
dipole cross sections
some hard diffractive processes at
moderately small Bjorken $x_B$
are unambigously calculable in terms of QCD evolution equation.
\par In the same time a detailed study of the space-time evolution of both
classical and quantum dipole in the center of mass-reference frame
of a dipole was undertaken by Gribov in ref. \cite{Gribov}, There
the space-time evolution of both the current and the
electromagnetic field was considered in the leading logarithmic
approach for zero mass particles. It was argued that such system
is unstable in the sense that the charges in the dipole are
screened out, and both the fields, the charges and the currents
flow to zero as $1/\log(t/\tau )$ for large times ($\tau$ is the
regularization parameter). Although the actual calculations were
carried for zero constituents masses it was argued that the
results are valid for small times as compared to the Compton waves
of the constituents.
\par More recently, a space-time evolution of dipoles was analyzed
in refs. \cite{Farrar}, \cite{FS1}. In particular, it was argued
in ref. \cite{Farrar}, that the behaviour of the dipole wave
function $\Psi(\rho,t)\sim \exp(i\rho^2/t))$, where $\rho$ is the
transverse scale of the dipole, leads to a new physical effect of
quantum diffusion, i.e. the average transverse area of a dipole
increases as $T/E$.
where $T$ is the time interval after dipole production.
   \par  In ref. \cite{FS1} a factorization theorem was proved for
the hard processes initiated by
longitudinal vector current, between the dipole
creation amplitude and the dipole scattering amplitude.
\par
Nevertheless the above approaches did not deal with a dipole as an
independent quantum object. Rather, a dipole was considered as a
product of two plane waves created by a local current. The total
gauge invariant amplitude including the dipole creation by a local
current and the corresponding scattering on the target was
considered using Feynman diagrams. In such approach dipole arises
as an intermediate state in Feynman diagram calculations and one
does not need to worry about its properties.

 The aim of the
present paper is to study whether the dipole in QED can be
described self-consistently, on the same basis as the point
-like
particle.
\par We shall consider approximately symmetric dipole, i.e. each of its
constituents carry approximately half of it's energy. We shall see
that such a dipole can be indeed described as a gauge invariant
way at least in the approximation of quantum mechanics,i.e. if we
neglect the radiation (particle production) and consider only
scattering. The parameters of the dipole are part of the initial
conditions one must specify, and are determined by the dipole
production process by a local current. We shall see that for times
small with the dipole coherence time, the expansion law of the
dipole is \beq
 \rho^2=DT/E+v_t^2T^2
\label{l0} \eeq Here $D$ is the diffusion coefficient and
$v_t=p_t/E$ is the average relative transverse velocity of the
constituents. In the leading order in T and for D=2, this equation
corresponds to the dipole created by the longitudinal part of the
vector current, coinciding with the result of ref. \cite{Farrar}.
\par The paper is organized in the following way. In the second
section we shall list the problems that appear when one tries to
describe the dipole in quantum field theory, problem of causality
and problem of gauge invariance and how they can be solved. In the
third chapter we study the time evolution of the dipole wave
function and a space-time evolution of a general free dipole in
quantum mechanics. In the fourth chapter we explain the need of
the regularization and study a space-time evolution in quantum
field theory (assuming the dipole is approximately symmetric, and
the characteristic transverse momenta of the constituents are much
smaller than the dipole center of mass energy). Our results are
summarized in conclusion. In Appendix we review for completeness a
classical analogue of a dipole creation and space-time evolution.
\par For
the sake of
simplicity throughout the paper we shall consider an
axially symmetric dipole with spinless constituents. We carry all our
calculations for the simpler QED case, although we expect that they will
remain true also for QCD.
\section{Problems with dipole in QED.}
 \subsection{Problem of casuality}
 \par We shall consider the fast free bare dipole
created at some time which we shall choose as $T=0$ by a  neutral
current. The word "bare" means here that a dipole created in this
way has no classical electromagnetic field. The  field is
regenerated, as in refs. \cite{Chudakov,Feinberg,Gribov}, starting
from $T=0$, and approaches a particular limit at
$T\rightarrow\infty$ (see Appendix for a review).
 The latter field will be different from the dipole analogue
of the Lienart-Wiehart potential, if we shall take into account
the dipole space-time evolution. Indeed, due to causality, being
zero at $T=0$, the classical field for each finite time $t$ will
be inside the sphere $r\le t$, and will not be spread to infinity
as  the field corresponding to the Lienart-Wiehart potential
\cite{LL}.
\par Consider the dipole created in this way. There are two
problems with its description in QFT. The first problem is
connected with casuality. Any process involving dipole, created at
T=0, must include vacuum fluctuations that existed since
$T=-\infty$. Thus in general all the boundary conditions on the
fields must be imposed at $T=-\infty$, and then of course there is
no sense to study separately the space-time evolution of the
dipole. However, it is easy to  see from the analysis of Feynman
diagrams along the lines of ref. \cite{Gribov1}  that since
perturbative QED vacuum is a ground state there exists a natural
time ordering in high energy dipole, with the processes involving
vacuum pairs, except properly ordered, being suppressed by the
powers of the energy (see also ref. \cite{Fubini-Furlan-Rosetti}).
This result permits one to use effectively old time-dependent
perturbation theory, and define the boundary conditions  at T=0,
and not at $T=-\infty$, that is the usual case in quantum field
theory \cite{Schweber}. The advantage of this approach is that it
enables one to take into account directly the space-time evolution
of the dipole wave function.
\par The direct calculation of Feynman diagrams shows that the sum of  the
terms unsuppressed by the powers of the center of mass energy of
the dipole $E$, exactly coincides with the result of the old
time-dependent perturbation theory (see e.g. refs. \cite{Grandy},
for a detailed set of rules for this formalism), with only one
causal ordering in time left. All other orderings are suppressed
due to energy denominators by the orders of energy E.
\par Note that in this case the old perturbation theory gives the
same results as the light-cone perturbation theory (see e.g. refs.
\cite{bjorken-soper,Brodsky} for the detailed review). Indeed, the
light-cone perturbation theory is causal in light-cone time $x^+$,
i.e. only causal light cone time ordering survives. It is clear
that for high energies time $t\sim z$, so $x^-\sim 0$, and $t\sim
x^+$. While for small energies the calculations are usually quite
complicated, for high energies the longitudinal and transverse
coordinates factorize, and the calculations of the scattering
processes in light-cone perturbation theory become quite simple.
However, there are two basic problems connected with the
systematic use of the light-cone gauge. First, contributions from
other time orderings really do not disappear, e.g. in QED
including fermions, but become contact terms that appear in loop
calculations, and are often hard to trace. Second, and more
important, the evolution of the wave functions that are solutions
of the wave equation, goes in time, note in light-cone time. One
has to specify boundary condition on the surface $t=0$, not
$x^+=0$. If one specifies initial conditions at $x^+=0$, they must
satisfy self-consistency conditions. This is of course not
surprising, since the solution of the wave equation depends on two
arbitrary functions, being the second order hyperbolic equation,
while the wave equation on light-cone is formally of the first
order in light-cone time $x^+$. In other words, not all boundary
conditions are allowed on light-cone, and they must be determined
from the dynamics of the theory. The same situation exists in
quantum field theory, where the boundary conditions at $x^+=0$
must be determined dynamically, even for high energies. Thus, once
the evolution of the wave functions is taken into account, one
either has to determine dynamically the evolution in $x+$ from
that in $t$ and then use light-cone perturbation theory, or
directly use the time-ordered perturbation theory as in ref.
\cite{Gribov1}. Consequently, the use of old perturbation theory
has its advantage since it permits one to consider boundary
problems with boundary conditions specified at T=0. The results
using both of these methods must be of course the same. Moreover,
technically the calculations in both approaches look very similar
at high energies. In our paper we shall use the old perturbation
theory, discarding the terms suppressed by powers of E.
\subsection{Problem of gauge invariance}
\par The problem of casuality described above actually exists also for
the evaluation of the e.m. field of a single particle. The problem
special for a dipole as an extended object is a problem of gauge
invariance of the calculated amplitudes.
\par It is well known (see ref. \cite{Schwinger}, and  ref. \cite{FS1}
for a more recent work on the subject) that in order for gauge
invariance to be preserved a dipole must be created by a local
current. It is easy to see that this condition is indeed necessary
and sufficient to ensure the gauge invariance if we neglect
radiation (i.e. particle creation).
 Indeed, consider the amplitude of the scattering of the
dipole on the external field in the lowest order over e.m. charge
e. \beq M_{if} =\lim_{T\rightarrow \infty} \int^T_0 dt d^3x
J_{(if)\mu}(\vec x,t) A_\mu (\vec x,t).\label{l1}\eeq If we make a
gauge transformation, the current is gauge invariant,
\cite{Schwinger},
 while the field changes as
$$A_\mu\rightarrow A_\mu+\partial_\mu \psi .$$
Here $\psi$ is the arbitrary function of $\vec x$ and t. Under
gauge transformation the amplitude $M_{if}$ changes as \beq \Delta
M_{fi}=\int dt\int d^3x J_\mu\partial^\mu \psi \label{l2}\eeq The
latter integral can be integrated by parts, and if as usual the
integration over t is from $- \infty$ to $+\infty$, and $\psi
(t=\pm \infty)=0$, the integral is \beq \Delta M_{if}=\int
d^4x\psi \partial^\mu J_\mu .\label{l3} \eeq This integral then
vanishes due to a current conservation. However because the
integration in eq. (\ref{l2}) is from $0$ to infinity, one obtains
additional term \beq \Delta M_{fi}=\int \rho_{fi}(\vec x,0)f(\vec
x)d^3x.\label{l4} \eeq Here $f(\vec x)=\psi(\vec x,0)$. Hence we
see that the condition for gauge invariance of the matrix elements
is that \beq \rho_{if}(\vec x)=0\label{l5}\eeq I.e. the density
matrix element for the transition between the initial state and
any arbitrary state f must be zero.
\par It is clear that the latter condition will be fulfilled if a dipole
is created by a local current. The proof above is not sufficient
for the case when we take into account radiation at finite time,
since then the operator of interaction may contain derivatives.
Taking radiation into account will be important for nonsymmetric
dipole and demands additional analysis \cite{Frankfurt1}.
\subsection{Gauge invariant initial conditions}
\par In the previous subsection we found the condition eq.
(\ref{l5}) for the initial dipole to be gauge invariant. It is
easy to see that for the initial wave function $\Psi$ this
condition is equivalent to \beq \Psi (\vec p+\vec k_1,\vec k_2)=
\Psi (\vec k_1,\vec p+\vec k_2)\label{l81} \eeq for arbitrary
$\vec p$.
 Suppose the initial wave
function can be separated  into
$$\Psi(\vec r_1,\vec r_2)=f_1(\vec r_1+\vec r_2)f_2(\vec r_1-\vec
r_2)$$ This separation corresponds to the separation of the center
of mass motion. Then it is easy to see that $f_2(\vec k)$ must be
independent of k, i.e. a constant.
\par Suppose for example
that the initial wave packet is axially symmetric and the
particles have no spin. Then we can use in the $\vec k$ space a
gaussian: \beq \lim_{a,\epsilon\rightarrow 0}
(2\epsilon/\pi)^{1/4}(2a/\pi)^{1/2}\exp (-aq^2_t-\epsilon q_z^2)
\label{l8}\eeq Such gaussian describes the initial axially
symmetric dipole, with the initial internal wave function $f_2$
normalized to 1, and the initial charge density is $\rho(\vec
x)=\delta (x)-\delta (\vec x)=0$. In general it is clear that one
can use as an initial wave function any function whose square is a
delta-function $\delta (\vec r_1-\vec r_2)$. In order to take into
account spin and angular moment one can either use the spherical
harmonics or non symmetrical regularization.
\section{Time evolution of the wave function of a neutral dipole.}
\subsection{General theory.}
\par  The key to the time-dependent description of the dipole in
quantum mechanics is its wave function. This wave function is a
solution of a Schroedinger equation:
  \beq \frac{\partial \Psi (\vec r_1,\vec r_2,t)}{\partial
t}=\hat H(\vec r_1,\vec r_2,t) \Psi (\vec r_1,\vec r_2,t)
\label{231} \eeq The Hamiltonian is \beq H=H_1+H_2-V(\vec r_1,\vec
r_2) \label{232} \eeq Here \beq H_i^2=(\vec p_i-e\vec A_i)^2
\label{233} \eeq In eq. (\ref{233}) $p_i$ is the momentum operator
for the i-th component, e is the charge of the constituent, V is
the potential of interaction between components of the dipole
(including the possible interaction with the external field) and
$A_i$ is the the quantum vector potential in the point $\vec r_i$,
determined from  the current and charge distribution due to the
wave function $\Phi$. In our approximation (no radiation) $A$ can
be considered as a classical field. \par The latter equation can
not be solved analytically in general, and here we shall consider
the case of the free dipole. In this section we shall consider a
free dipole without regularization. However, it is easy to see
that the results remain true also for interacting dipole, for
times much smaller than the coherence time.
 \subsection{Free dipole evolution.}
In this case we neglect the e.m. field acting on the components of
the dipole. Then the solution of eq. (\ref{232}) can be easily
written using momentum representation:
\begin{eqnarray}
\Phi(\vec r_1,\vec r_2,t)&=&\int
d^3q_1d^3q_2\Phi(\vec
q_1,\vec q_2)\exp(i\vec q_1\vec r_1+i\vec q_2\vec r_2)\exp(i\epsilon_{q_1}
+i\epsilon_{q_2})t\nonumber\\[10pt]
&=&\int d^3q_1d^3q_2\Phi(\vec Q,\vec q)\exp(i(\vec Q\vec R+\vec
q\vec r))\exp(i(\epsilon_{(Q+q)/2}
+i\epsilon_{(Q-q)/2})t)\nonumber\\[10pt]
\label{kj1}
\end{eqnarray}
Here $\epsilon_q=\sqrt{q^2+M^2}$, $\vec Q=\vec q_1+\vec q_2, \vec
R=(\vec r_1+\vec r_2)/2, \vec q=\vec q_1-\vec q_2, \vec r=(\vec
r_1-\vec r_2)/2$. This wave function is fully defined by its value
at t=0,i.e. by the boundary conditions \beq \Phi(\vec r_1,\vec
r_2,0)=\int d^3q_1d^3q_2\Phi(\vec q_1,\vec q_2)\exp(i\vec q_1\vec
r_1+i\vec q_2\vec r_2) \label{kj2} \eeq Depending on different
type of approximation we obtain different types of the space-time
evolution of the wave function.
\par First simplification is the "frozen dipole" approximation.
This approximation means that one assumes factorization between
the center of mass motion and the internal wave function, and the
internal wave function is assumed to be time independent. It is
clear from eq. (\ref{kj1}) that this approximation means that in
the time dependent energy multipliers we assume $\vert \vec Q\pm
\vec q\vert\sim \vert\vec Q\vert$,i.e. $\epsilon_{q_1}
+i\epsilon_{q_2}\sim 2\epsilon_Q\equiv E$ \beq \Phi(\vec Q,\vec
q)=\phi(q)/\sqrt{2(Q^2+M^2)}. \label{kj3} \eeq In this
approximation we obtain the wave function of the frozen dipole:
\beq \Phi(\vec R,\vec r,t)=\frac{\exp(i\vec Q\vec
R)}{\sqrt{2\epsilon_Q}}f(\vec r)\exp (iEt). \label{kj4} \eeq The
latter wave function describes the evolution of the dipole with
the energy E, c.m. momentum $\vec Q$ and without internal
dynamics.
\par The simplest internal wave function that can be taken is the Gaussian
wave packet in the transverse plane and a narrow slap in the
z-direction that approximately corresponds to the delta-function
due to the relativistic effects: \beq
f(q)=(2a/\pi)^{1/2}(2\epsilon/\pi)^{1/4}\exp(-(a+ib)(q_t-k_t)^2)\exp(-\epsilon
q_z^2). \label{kj5}\eeq Here $\vec \rho =\vec \rho_1-\vec \rho_2$,
 z is the direction of the motion of the dipole,  $\rho$ is
the transverse coordinate, and $\vec k_t$ is the transverse
momenta of the constituents in the transverse plane.
 Note that a general initial wave function is
complex. The limit $ \epsilon\rightarrow 0$ corresponds to the
delta function distribution of the matter in the relativistic
disc. Index t means the direction transverse to the direction of
$\vec Q$. In the coordinate space we obtain \beq f(\vec \rho ,
z_1,z_2)=\frac{\sqrt{2a\pi}}{a+ib}
\exp(-\rho^2/(4(a+ib)))(2\pi/\epsilon)^{1/4}
\exp(-(z_1-z_2)^2/(4\epsilon))\exp(i\vec k_t\vec \rho).
\label{jkl6} \eeq For the state described by eq. (\ref{jkl6})

 \begin{eqnarray}
<\vec
p^i_t(t=0)>&=&k^i_t, <p^2(t=0)>=2/a+k^2_t\nonumber\\[10pt]
<r^i(t=0)>&=&0,<r^2(t=0)>=(a+b^2/a)/2.\nonumber\\[10pt] \label{27}
\end{eqnarray}
\par The frozen dipole approximation is useful because we do not have to
think about wave function dynamics and can put the boundary conditions
at $t=-\infty$.
\par We can approximately take into account the evolution of the free dipole
explicitly for small times. The simplest way to do it is,
neglecting for simplicity the masses of the constituents,  to
expand \beq \vert \vec Q\pm \vec q\vert=Q\pm q_z+q^2_t/(2Q).
\label{kj8} \eeq Then each of the time dependent multipliers in
the wave function (\ref{kj1}) can be written as
$$\exp (i(Et/2\pm q_z+q^2_t/(E))t)$$
The gaussian wave packet (\ref{kj5}) in such  approximation
acquires a time dependence
\begin{eqnarray}
\Phi(\vec r,\vec R,t)&=&(1/\sqrt{2E})\sqrt{2a/\pi}\exp(i\vec Q\vec
R)\exp(iEt)\int
d^2q_t\exp(-(a+ib)(\vec q_t-\vec k_t)^2+2(it/(E))q^2_t)\nonumber\\[10pt]
&\times&\exp(i\vec q_t\vec \rho ),\nonumber\\[10pt]
\label{kj9}
\end{eqnarray}
(where we neglect the frozen evolution in z direction). In the
coordinate space one obtains in this approximation
\begin{eqnarray}
\Phi(r,R,t)&=&(1/\sqrt{2E})\exp(i\vec Q\vec R)\exp(iEt)
\frac{\sqrt{2a\pi}}{a+i(b+4t/E)}\nonumber\\[10pt]
&\times&\exp{(-(\vec \rho -4\vec k_tt/E)^2/((a+i(b+4t/E))+i\vec
k_t(\vec \rho-4\vec k_tt/E)}). \nonumber\\[10pt]
 \label{34} \end{eqnarray}
 For this dipole we
have time dependent radius: \beq \rho^2(t)=a/2+b^2/2a+8(b/a)\hbar
t/E+8t^2/(E^2a) \label{j34} \eeq \beq \rho^i(t)=4k^it/E.
\label{j35} \eeq
\par We  wrote $\hbar$ in the latter equation in order to emphasise that
the linear  in time term has purely quantum origin. Let us note
that this term is nonzero only for  complex initial conditions.
This term  corresponds to the quantum diffusion phenomena
\cite{Farrar}. The term in the expansion $\sim t^2$ corresponds to
the usual classical dynamics, this is the only term surviving for
the real wave packet. The wave function (\ref{34}) corresponds to
the dipole built of the two components of opposite charges that
are separated in the transverse plane by the vector $\vec \rho
(t)=4\vec k_tt/E$. For $\vec k_t\ne 0$ such a dipole has a well
defined classical limit, describing two constituents, each with
the transverse velocity $\vec v_t=\vec k_t/(E/2)$, and we shall
call such a dipole a quasiclassical dipole. However if $\vec
k_t=0$, the dipole classical limit is not well defined. The dipole
can't be considered as being built of two separate components, and
is locally electrically neutral. We shall call below such object a
quantum dipole. I.e. by definition a quantum dipole is a dipole
such that by definition \beq <\vec r_1(t)-\vec r_2(t)>=0.
\label{lkj1} \eeq
\par Below we shall need a limit of a very small dipole:
 \beq b/a\rightarrow 1, a\rightarrow 0, b\rightarrow 0. \label{35} \eeq
\par It is easy to check that for small times, i.e. times much
less than the coherence time $T_c\sim E/p^2_t$, where $p_t$ is the
characteristic transverse momenta in the dipole, the interactions
do not change qualitatively the dipole dynamics. Moreover, the
quantum diffusion term is not influenced by the interaction
potential $V$, and remains the same as for the free particle.

 \section{Regularization of
a free neutral dipole.} \par In the previous section we considered
the time evolution of the free dipole in quantum mechanics. In
order to go to QED, and to consider dipole created by a local
current, one needs, as it was explained in section 2, to take the
limit $a,b\rightarrow 0$ ( in the notations of the previous
section). If one takes this limit in the space-time wave function,
one obtains the wave function \beq f(\vec \rho,t)\sim
\exp(i\rho^2E/t).\label{l10}\eeq This wave function however can
not be normalized to one, since the preexponent does not have a
well defined limit. Note that this wave function naively implies
quantum diffusion \cite{Farrar}: Although $<\rho^2>$ is not well
defined, the phase of this wave function strongly oscillates at
distances $\rho^2\sim t/E$. \par It is clear why there are
problems with the $a,b\rightarrow 0$ limit. Indeed, $1/\sqrt{a}$
is, according to the uncertainty principle, average momentum of
transverse quantum fluctuations inside a dipole. When $a$ tends to
zero, we eventually come to the situation when $1/\sqrt{a}/E\sim
1$,i.e. there is no separation between transverse and longitudinal
degrees of freedom for any sensible period of time. The motion of
transverse degrees of freedom is relativistic.  It is easy to see
that the same problem arises if one does not carry the separation
of variables as in  the previous section, but considers the
evolution of the ultrarelativistic dipole from the beginning using
the saddle point method. \par The only way to solve the problem is
 to take the limits  $E\rightarrow\infty$ and $a\rightarrow 0$
simultaneously. Let us look at the expression for $<\rho^2>$. It
is clear that physically the term proportional to $t^2$ for free
dipole is just $v^2_tt^2=p_t^2/E^2t^2$. This means that for
quantum dipole ($k_t$ in the previous section equal to zero), one
must take a limit $E\rightarrow \infty$, $a\rightarrow 0$,
$E^2a\rightarrow 1/v^2_t$. Thus we see that in order to define the
evolution of quantum dipole, in addition to an initial wave
function and center of mass energies of the components one must
define the way how the wave function and physical quantities are
regularised. The regularisation parameters such as $v_t$ are
additional parameters for a quantum dipole that must be specified
as the initial conditions, in addition to the initial wave
function. Physically they are determined by the physical process
that creates dipole, and may be different for different local
currents. In order to determine them we must consider in detail
the process of the dipole creation in QED.
 \par We now turn to quantum diffusion.
It is clear that in the $a,b\rightarrow 0$ limit we must also
demand $b^2/a\rightarrow 0$. It is also clear we must demand that
$8b/a\rightarrow {\rm const} \equiv D$. This means that b and a
tend to zero with the same speed. The quantum diffusion
coefficient D must be specified as an initial condition. As it was
explained above, its value depends on the explicit form of the
current creating the dipole. For example, for the case of the
longitudinal part of the vector current, such analysis gives  D=2
\cite{Farrar,Doc}.  \par Above we considered the case of the
quantum dipole, i.e. $\vec k_t=0$ in the notations of the previous
section. Such dipole is locally charge neutral even for finite
times, and does not have well-defined quantum limit. (The average
$<p^2_t>=v^2_tE^2$ for such a dipole of course can be nonzero).
 If $k_t\ne 0$, the only change with the previous
analysis is that the classical transverse separation speed $v_t$
is now a sum \beq v^2_t=u^2_t+p_t^2/E^2.\label{p12}\eeq For a
space-time evolution of the dipole in quantum field theory one
obtains \beq \rho^2=Dt/E+v_t^2t^2\label{h12}\eeq Other physical
quantities must be renormalized along the same lines. Note that
our approach to regularization looks very similar to the general
Schwinger regularization \cite{Schwinger}.
 \section{Conclusion.} \par In this paper we considered the
possibility of description of a space-time evolution of a quantum
dipole in QED. We have considered a dipole created at the time T=0
by a local current. The dipole was assumed to be approximately
symmetric (i.e. each constituent carries approximately half of its
energy), with the center of mass energy $E$ much higher than the
characteristic momenta of the transverse degrees of freedom. We
have found that such  dipole can be described in a gauge invariant
way by the time dependent wave function, at least if one neglects
radiation at finite time $T>0$.
\par We have
seen that  the condition of gauge invariance of scattering
amplitude (excluding radiation) strongly restricts the possible
form of the dipole initial wave function that must be localized in
a single point. Moreover, to specify the initial conditions it is
not enough to specify  space coordinate dependence of the initial
wave function, as it will seem naively. In addition one must
specify the renormalization procedure and the values for the
renormalized physical parameters, the quantum diffusion
coefficient D and the average transverse velocity. The values for
these parameters are determined by the dipole creation process and
depend on the explicit form of the current creating a dipole.
\par We were able to give the exact definition to the concepts of
the space-time evolution of the quantum dipole, quantum diffusion
, and an average size of the quantum dipole as a renormalized
quantity. This concepts were first discussed on a more intuitive
level in ref. \cite{Farrar}.
\par In this paper we considered a symmetric dipole. It is easy to
see that along the same lines one can consider asymmetric dipole,
such that the dipole constituents have the energies $xE $ and
$(1-x)E$. The only difference will be that in the equations for
the dipole size and the wave functions E must be substituted by
$Ex(1-x)$. However, if $x$ is significantly different from 1/2 the
radiation becomes important and the additional analysis of the
dipole evolution is needed \cite{Frankfurt1}.
\par Another interesting question is the dipole evolution in the external
field how to take into account  the interaction between the
constituents of the dipole. This is  a subject for the future
studies, although we expect it will not change qualitatively the
results obtained using a free dipole. It is clear that in this
case we must use the full system of the wave functions in the
external field. It will be very interesting also to study the
effects of the spin on the dipole evolution.
\par Next, it will be interesting to study how and for what times
bound states influence the dipole space-time evolution.
\par Finally, it will be very interesting to include the photon radiation
in the above formalism. \acknowledgements{The author thanks
Professor L. Frankfurt for reading a paper and numerous useful
discussions.}
 \appendix
\section{}
\subsection{Field generation by a scalar particle.}
\par In this appendix we discuss the issue of field generation by
a classical dipole, consisting of two distinct components
spreading with a definite average velocity. Similar, but more
complicated theory of a field generation may be developed for a
quantum dipole.
\par We start by showing what we mean by field generation by bare particle.
We shall start from a purely mathematical problem of solving the
wave equation with the source appearing at time t=0 and moving
rapidly. Similar problem of the charged particle field change
during sudden stop or acceleration was considered in refs.
\cite{Feinberg},\cite{Bolotovskii}.
 The wave
equation has the form: \begin{eqnarray}
\Box \phi (\vec x,t)&=&\rho (\vec x,t)\nonumber\\[10pt]
\rho (\vec x,t)&=&\delta (z-vt)\delta (\vec \rho )\theta (t)\nonumber\\[10pt]
\label{1}
\end{eqnarray}
Note here the appearance of theta function meaning that the source
appeared at $t=0$. The speed $v$ is assumed close to the velocity
of light $v\sim 1$.
\par The eq. (\ref{1}) can be easily solved by making Fourie transform both
in $\vec x $ and in  t: \beq \phi (\vec q,\omega)=\int d^3\vec xdt
\phi(\vec x,t)\exp (-i\vec q\vec x-i\omega t) \label{2} \eeq
Making Fourie transform of both parts of eq. (\ref{2}) we obtain
\beq \phi(\omega, \vec q)=-\frac{1}{\omega^2-q^2+i\epsilon {\rm
sign}\omega} \frac{1}{\omega-\vec q\vec v-i\epsilon}. \label{3}
\eeq Note the prescription for pole we made. This prescription of
course corresponds to using the retarded Green function while
solving eq. (\ref{1}), so that all the poles are in the upper
half-plane. Then the solution is zero for $t\le 0$. In order to
understand the form of the solution in the space-time we shall do
the inverse Fourie transform. First, let us do the transform in
the $\omega$ space. We have to sum over 3 poles:$\omega =\pm q
+i\epsilon,\omega =\vec q\vec v+i\epsilon$. Note that the third
pole is just the Lienart-Wiehart contribution, while the first two
poles correspond to on-shell field. Explicitly, we obtain: \beq
\phi (\vec q,t)=-\frac{\cos (qt)-\cos (\vec q\vec
vt)}{(q^2_t+(1-v^2)q_z^2)}-i \frac{q\sin(qt)-\vec q\vec v\sin(\vec
q\vec vt)}{q(q^2-(\vec q\vec v)^2}). \label{4} \eeq
\par
It is worthwhile to go further and do Fourie transform in q
\cite{BE}. We see that the Fourie transform of the first term in
$\vec q\vec v$ is different from zero only for $r\ge t$, and for
$r\ge t$ the first and the second term in eq. (\ref{4}) exactly
cancel each other, while for $t\ge r$ only the second term
contributes and gives exactly the Lienart-Wiehart contribution
(or, more explicitly the contribution of the field of rapidly
moving scalar charge).Thus we come to a solution \beq \phi (\vec
x,t)=\theta (t-r)\phi_0(\vec x,t). \label{5} \eeq Here \beq
\phi_0(\vec x,t)= \frac{1}{\sqrt{ \rho^2(1-v^2)+(z-vt)^2}}.
\label{6} \eeq This is  the field of the rapidly moving scalar
source \cite{LL} created at the moment of time t=0. The theta
function imposes casuality (The details of the Fourie transform
are given in the next subsection). \subsection{Details of the
Fourie transform of the potentials.} In order to find explicitly
the space structure of the solutions discussed above, the easiest
way is to use explicitly the retarded Green function. Using  the
Fourie transform  \beq G_{\rm ret}(\omega, \vec
q)=1/(\omega^2-q^2+i\epsilon {\rm sign}\omega ),\label{A1} \eeq
and \cite{BE} \beq G_{\rm ret}=\delta (r-t)/r \label{A2} \eeq and
taking into account that under Fourie transform the product
becomes a convolution, we obtain the field $\phi$ : \beq \phi
(\vec r,t)=\int d\tau d^3 r^{'} \frac{\delta (t-\tau-\vert \vec
r-\vec r^{'}\vert )}{\vert \vec r -\vec r^{'}\vert }
\rho(r^{'},\tau) \label{A3} \eeq Substituting the expression for
density (\ref{3}): $\delta (z'-v\tau)\delta(\rho')$, we
immediately obtain: \beq \phi( r,t)=\int^t_0d\tau
 \frac{\delta (t-\tau-\sqrt{\rho^2+ (z-v\tau)^2)}}{\sqrt{\rho^2+(z-v\tau
)^2)}}. \label{A4} \eeq It is straightforward to check by solving
a quartic equation that the argument of delta function has a zero,
such that $t\ge \tau\ge 0$ only if $t\ge r$, and vice versa. Thus
we come to a space-time structure for a solution for $\phi$ in the
previous subsection: \beq \phi (\vec r,t)=\theta
(r-t)\frac{1}{\sqrt{\rho^2 (1-v^2)+(z-vt)^2}}. \label{A5} \eeq
 \subsection{Expanding dipole.}
\par The results of the previous subsection can be extended to
electrodynamics, the only subtlety is that we have to ensure the
conservation of the electric current: \beq \frac{d\rho}{dt}+{\rm
div}\vec j=0 \label{7}. \eeq The case of the field change for
sudden stop or acceleration of the charge was considered in ref.
\cite{Bolotovskii}. Here we consider a case of a creation of a
point-like classical dipole that  satisfies eq. (\ref{7}). Suppose
the two components of the dipole were created at T=0, in such a
way that they both have a speed component v along z axis and equal
but opposite in magnitude velocities in the transverse plane $\vec
v_t$ and $-\vec v_t$. Such symmetric dipole is of course over
simplification in real laboratory system, where target is at rest,
but the general case is not much different. The charge and current
density for such classical symmetric dipole are given by
\begin{eqnarray}
\rho (\vec r,t)&=&\delta (z-vt)*(\delta (\vec \rho-\vec
v_tt)-\delta (\vec\rho+\vec v_t t))\theta (t)\equiv \rho_+(\vec
r,t)-\rho_-(\vec
r,t)\nonumber\\[10pt]
 j_z(\vec r,t)&=&v\rho (\vec r,t)\nonumber\\[10pt]
\vec j_t(\vec r,t)&=&\vec v_t( \rho_+(\vec r,t)+\rho_-(\vec
r,t)).\nonumber\\[10pt]
\label{8}
\end{eqnarray}
It is straightforward to check that this ansats satisfies charge
conservation equation (\ref{7}). In writing eqs. \ref{8} we
neglected the further evolution of the dipole due to coulombic
interaction between the constituents.
 The Maxwell equations can be easily solved both in
Lorentz and in Colulomb gauges.
 \begin{eqnarray}
\triangle
A_\mu-\displaystyle{\frac{d^2A_\mu}{dt^2}}&=&-j_\mu\nonumber\\[10pt]
\displaystyle{\frac{\partial^i A_i}{\partial x^i}+dA_0/dt}&=&0\nonumber\\[10pt]
\label{911}
\end{eqnarray}
Making a Fourie transform as in the previous section we obtain the
solution in Lorentz gauge:
\begin{eqnarray}
A_0=i(\phi_1-\phi_2),\,\,\,\,A_3=vA_0\nonumber\\[10pt]
 \vec A_t^i=iv^i_t(\phi_1+\phi_2)\nonumber\\[10pt] \label{10}
\end{eqnarray}
Here $\phi_i$ are given by eq. (\ref{3})  for $(\omega,\vec q)$
and (\ref{4}) for $(\vec q,t)$ representations, with $v=v_1$ for
$\phi_1$ and $v=v_2$ for $\phi_2$. In our kinematics $\vec
v_1=(v,\vec v_t),\vec  v_2=(v,-\vec v_t)$.
\par The same solution in Coulomb gauge is
\begin{eqnarray}
\phi=\phi_1 (1-\omega (v_1k)/k^2)-\phi_2(1-\omega (\vec v_2\vec
k)/k^2))\nonumber\\[10pt]
\vec A_i=(\vec v_1-\vec k(v_1k)/k^2)\phi_1-(\vec v_2-\vec k (\vec
v_2\vec k)/k^2)\phi_2. \nonumber\\[10pt]
\label{10a}
\end{eqnarray}

 Once again all poles are in the upper
half plane to ensure that the solution is zero at $t<0$. It is
clear that the space-time structure of the solution will be the
same as in the previous chapter: it will be the standard dipole
field multiplied by a casuality ensuring theta-functions $\theta
(r_i-t)$. It is straightforward to move into mixed $q-t$
representation by making Fourie transform in $\omega$
\par Let us find the electric and magnetic field
corresponding to these vector potentials. We shall do it first in
$\omega , \vec q$ and in $\vec q,t$ spaces. In the Fourie
components we have:
\begin{eqnarray}
E^i_t=iq^i_t(\phi_1-\phi_2)-\omega v^i_t(\phi_1+\phi_2)\nonumber\\[10pt]
E_3=i(q^3v-\omega)(\phi_1-\phi_2)\nonumber\\[10pt]
H_{ti}=\epsilon_{i3k}iq^3v_{t2}(\phi_1+\phi_2)-i\epsilon_{ik3}q^kv(\phi_1-\phi_2)
\nonumber\\[10pt]
H_3=i\epsilon_{ij}q^iv_{tj}(\phi_1-\phi_2).\nonumber\\[10pt]
\label{13}
\end{eqnarray}
\par In the space-time representation  we have
\beq \vec E=\sum_{i=1,2}\vec E_1\theta (r_i-t)+\vec E_2\delta
(t-r_i)\,\, \vec H=\sum_{i=1,2}\vec H_1\theta (t-r_i)+\vec
H_2\delta (t-r_i) \label{13a}
\end{equation}
Here $\vec E_1,\vec H_1$ are the usual Lienart-Wiehart field of
the moving charge, however they now exist only inside the
light-cone due to causality. The field $E_2,H_2$ exists only on
light-cone. It is easy to see that the fields $E_2$ and $H_2$ are
orthogonal in space-time and to light cone, for each of the
constituents and thus correspond to the electromagnetic waves that
exists on the boundary of light cone-i.e. on the surface $r=t$.
The explicit form of these fields is easily restored using the
Fourie transforms of appendices A and B.

\end{document}